\documentclass[twocolumn,prd,aps,superscriptaddress,showpacs,preprintnumbers]{revtex4}

\usepackage{amsmath}
\usepackage{verbatim}
\usepackage{graphicx}
\usepackage[usenames]{color}
\usepackage{psfrag}
\usepackage{epstopdf}
\usepackage{hyperref}
\def\beq{\begin{equation}}
\def\eeq{\end{equation}}
\def\bea{\setlength\arraycolsep{1.4pt}\begin{eqnarray}}
\def\eea{\end{eqnarray}}
\def\bit{\begin{itemize}}
\def\eit{\end{itemize}}

\begin{document}
\preprint{\tt NPAC-09-XX}

\title{Constraints on the anisotropy of dark energy}
\author{Stephen Appleby} \email{sappleby@jb.man.ac.uk}
\affiliation{Jodrell Bank Center for Astrophysics, School of Physics and Astronomy, University of Manchester, Manchester, M13 9PL  UK}
\author{Richard Battye} \email{rbattye@jb.man.ac.uk}
\affiliation{Jodrell Bank Center for Astrophysics, School of Physics and Astronomy, University of Manchester, Manchester, M13 9PL  UK}
\author{Adam Moss} \email{adammoss@phas.ubc.ca}
\affiliation{ Department of Physics \& Astronomy, University of British Columbia, Vancouver, BC, V6T 1Z1  Canada}

\date{\today}

\begin{abstract}
If the equation of state of dark energy is anisotropic there will be additional quadrupole anisotropy in the cosmic microwave background induced by the time dependent anisotropic stress quantified in terms of $\Delta w$. Assuming that the entire amplitude of the observed quadrupole is due to this anisotropy, we conservatively impose a limit of $|\Delta w| < 2.1\times 10^{-4}$ for any value of $w\ge -1$ assuming that $\Omega_{\rm m }<0.5$. This is considerably tighter than that which comes from SNe. Stronger limits, upto a factor of 10,  are possible for specific values of $\Omega_{\rm m}$ and $w$. Since we assume this component is uncorrelated with the stochastic component from inflation, we find that both the expectation value and the sample variance are increased. There no improvement in the likelihood of an anomalously low quadrupole as suggested by previous work on an elliptical universe. 
\end{abstract}

\maketitle

{\em Introduction:} There is now very strong evidence that expansion of the Universe is accelerating based on a combination of measurements of type Ia supernovae (SNe)~\cite{Riess:1998cb,Perlmutter:1998np}, anisotropies of the cosmic microwave background (CMB)~\cite{Spergel:2006hy,Komatsu:2008hk} and large scale structure~\cite {Eisenstein:2005su}. Many different explanations for this phenomenon have been suggested with most of them requiring the existence of new physics at a scale comparable to the present horizon size $\sim c/H_0$ (for example, refs.~\cite{Copeland:2006wr,Sotiriou:2008rp}).  One of these ideas is that there is some new energy component, known as dark energy, with an equation of state $P=w\rho$ and $w<-1/3$ in order to achieve acceleration. 

Measurements of CMB anisotropies appear to show that there a number of anomalies in their detailed statistical properties~\cite{Eriksen:2003db,Land:2005ad,Jaffe:2005pw,Hoftuft:2009rq}. These suggest that the anisotropies may not be compatible with an isotropic, Gaussian random field on the very largest scales. Given that the Fourier modes corresponding to these anisotropies crossed the horizon in the last few Hubble times, that is, when the dark energy has come to dominate the expansion of the Universe, it is tempting to connect the two.

The equation of state parameter $w$ is often considered to be a function of time, but the overall pressure tensor, ${P_i}^{j}$, is usually assumed to be isotropic, that is ${P_{i}}^j=w\rho_{\rm de}{\delta_i}^{j}$ where $\rho_{\rm de}$ is the density of the dark energy.  In this {\it letter} we will consider the possibility that the dark energy is parameterized by
\begin{equation}
{P_i}^{j}=\rho_{\rm de} \left[ w {\delta_i}^{j}+{\Delta w_{i}}^{j} \right]\,,
\label{pressure}\end{equation}
where we impose the traceless condition ${\Delta w_i}^{i}=0$. We will show that an anisotropic equation of state leads to an additional component in the quadrupole anisotropy and hence the observed amplitude of the CMB quadrupole leads to a strong constraint on $\Delta w_{ij}$ which is much stronger than those which have been deduced from SNe data~\cite{Koivisto:2007bp,Koivisto:2008ig,Cooke:2009ws}.

The study of anisotropic universes has a long history~\cite{Hawking:1968zw,Collins:1972tf,Barrow}, but in the past the possibility that the Universe is rotating, being modelled by a non-trivial Bianchi universe, was the main focus. In our work we will concentrate on models which only have time dependent and spatially homogeneous anisotropic stress and no vorticity. Moreover, it is only the dark energy component which is anisotropic allowing the evolution of the Universe to proceed in the standard way until the dark energy begins to dominate.

An anisotropic equation of state is natural in the Elastic Dark Energy (EDE) model~\cite{Battye:2006mb,Battye:2007aa}, whereby the dark energy component is comprised of some continuum fluid with a non-zero shear modulus, similar to the idea of a crystalline material. One particular manifestation of this idea could be a soap film type structure formed from domain walls at a low energy phase transition with $w=-2/3$~\cite{Battye:1999eq}. However,  the limits which we will derive apply to any particular manifestation of anisotropic dark energy, for example, as discussed in refs.~\cite{Koivisto:2007bp,Koivisto:2008ig}

The reason for this is that, to first order, the anisotropic evolution due to ${\Delta w_{i}}^{j}$ and the evolution of  initial metric perturbations decouple and can be computed independently, then added at the end. If we use the time dependent spatial metric $\gamma_{ij}(\eta)$ to describe the  anisotropic evolution~\cite{Pereira:2007yy} and $h_{ij}(\eta,{\bf x})$ to represent effect the metric perturbations created during inflation, then our approach works when $|\gamma_{ij}|\gg |h_{ij}|$.  It should also yield qualitative information when $\gamma_{ij}\sim h_{ij}$.

{\em Anisotropic universes:} We will consider universes with metric 
\begin{equation}
ds^2=a^2\left(-d\eta^2+\gamma_{ij}(\eta)dx^idx^j\right)\,,
\end{equation}
for which ${\Gamma^{i}}_{0j}={\cal H}{\delta_i}^{j}+{\sigma_j}^{i}$ and ${\Gamma^{0}}_{ij}={\cal H}\delta_{ij}+\sigma_{ij}$, where $\sigma_{ij}={1\over 2}{d\over d\eta}\gamma_{ij}$ and ${\cal H}=a^{\prime}/a$, with the derivative with respect to conformal time, $\eta$. The stress-energy content will be made up of dark matter (m), which will be isotropic and pressureless,  and dark energy (de), where the pressure tensor of the dark energy is given by (\ref{pressure}).
The Einstein and conservation equations can be written as~\cite{Pereira:2007yy} 
\begin{eqnarray}
3{\cal H}^2&=&8\pi Ga^2\rho_{\rm tot}+{1\over 2}\sigma^2\,,\\
{\rho_{\rm de}}^{\prime}&=&-3{\cal H}(1+w)\rho_{\rm de}-{\sigma_{j}}^{i}{\Delta w_{i}}^{j}\rho_{\rm de}\,,\\
 {{{\sigma}_{i}}^{j}}^{\prime}&=&-2{\cal H}{{\sigma}_{i}}^{j}+8\pi Ga^2{\Delta w_{i}}^{j}\rho_{\rm de}\,,
\end{eqnarray}
where $\rho_{\rm tot}=\rho_{\rm m}+\rho_{\rm de}$, $\sigma^2={\sigma_i}^{j}{\sigma_j}^{i}$ and we have used a flat geometry. One might think that there would be a velocity generated , but this would only be created by the gradient of the anisotropic stress which is absent in our model. Hence, we can consistently set this to zero and ignore it. Assuming there is no primordial anisotropic stress, that is, initially ${\sigma_i}^j=0$, these equations can be solved to give $\rho_{\rm de}$ and ${\sigma_{i}}^{j}$ as functions of $a$ if ${\Delta w_i}^{j}$ is considered small. To order $(\Delta w)^3$, we find that 
\begin{eqnarray}
\rho_{\rm de}(a)&=&{\rho_{\rm de}(t_0)\over a^{3(1+w)}}\left(1-3\Omega_{\rm de}{\Delta w_i}^j{\Delta w_j}^iG(a)\right)\,,\\
{{\sigma_i}^j(a)\over H_0}&=&{3\Omega_{\rm de}{\Delta w_i}^{j}\over a^2}F(a)\,,
\end{eqnarray}
where $E(a)=(\Omega_m/a^3+\Omega_{\rm de}/a^{3(1+w)})^{1/2}$, 
\begin{eqnarray}
F(a)&=&\int_0^a{db\over b^{1+3w}E(b)}\,,\\
G(a)&=&\int_0^a{db\,F(b)\over b^4E(b)} \,,
\end{eqnarray}
and $\Omega_{\rm X}$ is the density of species X relative to the present day critical density, with $\Omega_{\rm m}+\Omega_{\rm de}=1$. The function $F(a)/a^2$ is the growth function for anisotropic stress.

\begin{figure}
\centering
\mbox{\resizebox{0.45\textwidth}{!}{\includegraphics[angle=0]{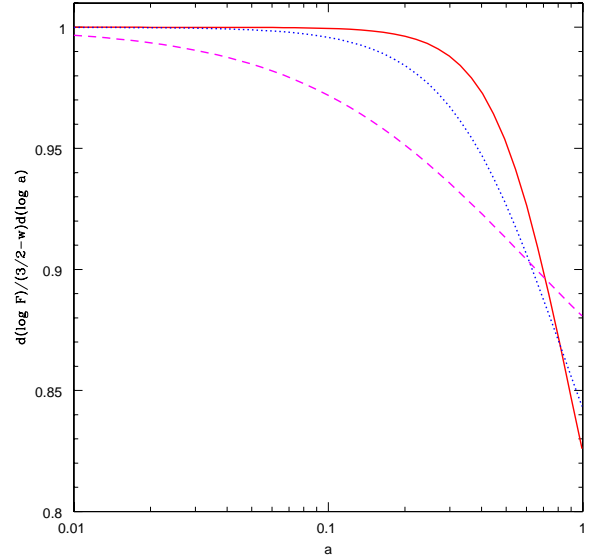}}}
\caption{\label{fig:fff} The power law slope of the function $F(a)$ for $w=-1$ (solid line), $-2/3$ (dotted line), $-1/3$ (dashed line). In each case we have used $\Omega_{\rm m}=0.3$.}
\end{figure}

The anisotropic stress grows quickly during matter domination with $F(a)\approx a^{(3/2)-3w}/(\Omega_{\rm m}^{1/2}((3/2)-3w))$ and the rate of increase declines during the dark energy dominated era. We have plotted $d(\log F)/(3/2-w)d(\log a)$ against $a$ for a range of values for $w=-1$, $-2/3$ and $-1/3$ for $\Omega_{\rm m}=0.3$ in Fig.~\ref{fig:fff}. The slopes start at 4.5, 3.5 and 2.5, respectively, and decrease with time; this happens later in models with more negative values of $w$, since dark energy domination happens later for fixed $\Omega_{\rm m}$.

\begin{figure}
\centering
\mbox{\resizebox{0.45\textwidth}{!}{\includegraphics[angle=0]{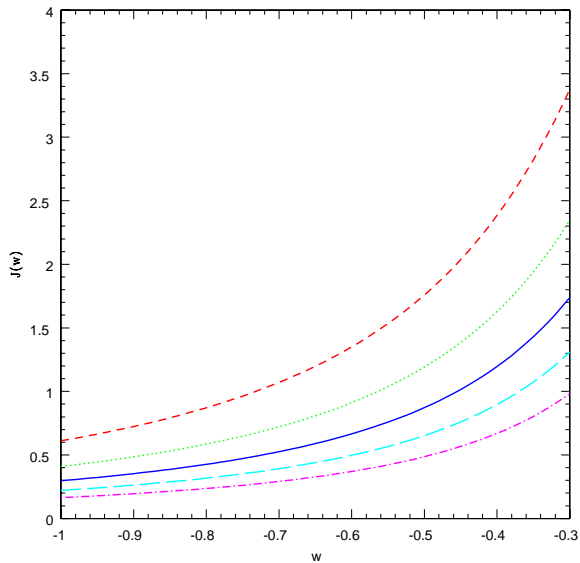}}}
\caption{\label{fig:jjj} The function $J(\Omega_{\rm m},w)$ against $w$ for $\Omega_{\rm m}=0.1$ (dashed line), 0.2 (dotted line), 0.3 (solid line), 0.4 (short-dashed line) and 0.5 (dotted-short dashed line). }
\end{figure}

{\em Calculation of temperature anisotropies:}  The CMB anisotropies due to the anisotropic equation of state are given by
\begin{equation}
{\Delta T\over T}({\hat n})=-\int_{\eta_{\rm rec}}^{\eta_0}\sigma_{ij}{\hat n}^{i}{\hat n}^{j}d\eta\,,
\end{equation}
where $\eta_0$ and $\eta_{\rm rec}$ are the conformal time at the present and recombination eras, respectively.
Using the expression above with $\sigma_{ij}=\gamma_{ik}{\sigma_j}^{k}$ and converting the integration variable to $a$ (ignoring terms higher order in $\Delta w$), one finds that 
\begin{equation}
{\Delta T\over T}({\hat n})=-\Delta w_{ij}{\hat n}^i{\hat n}^jJ(\Omega_{\rm m},w)\,,
\end{equation}
where 
\begin{equation}
J(\Omega_{\rm m},w)=3(1-\Omega_{\rm m})\int^{1}_{a_{\rm rec}}{da\over a^4E(a)}\int^a_0{db\over b^{1+3w}E(b)}\,,
\end{equation}
and $a_{\rm rec}\approx 1/1090$ is the scale factor at recombination. Calculation of $J(\Omega_{\rm m},w)$ will be insensitive to the precise value of $a_{\rm rec}$ used because it is dominated by the late time behaviour. We have plotted $J(\Omega_{\rm m},w)$ against $w$ for a range of values of $\Omega_{\rm m}$ in Fig.~\ref{fig:jjj}. It is clearly ${\cal O}(1)$ for the range of values relevant to observations.

It may appear that the anisotropic equation of state parameter, $\Delta w_{ij}$, has 5 degrees of freedom since it is a traceless, symmetric, rank 3 matrix. However, one can rotate the coordinate system of the observation, such that ${\hat n}$ can be replaced by $R{\hat n}$. This corresponds to a diagonalization of $\Delta w_{ij}$ and hence it is sufficient to parameterize $\Delta w_{ij}={\rm diag}(\Delta w_1,\Delta w_2,-(\Delta w_1+\Delta w_2))$ and the rotation $R$ which defines the direction of the anisotropy on the celestial sphere. The limits we will compute will not depend on $R$. Hence, we can write 
\begin{eqnarray}
{\Delta T\over T}({\hat n})=-J(\Omega_{\rm m},w)\bigg(\Delta w_1\sin^2\theta\cos^2\phi\nonumber\\
+\Delta w_2\sin^2\theta\sin^2\phi-(\Delta w_1+\Delta w_2)\cos^2\theta\bigg)\,.
\end{eqnarray}

\begin{figure}
\centering
\mbox{\resizebox{0.45\textwidth}{!}{\includegraphics[angle=0]{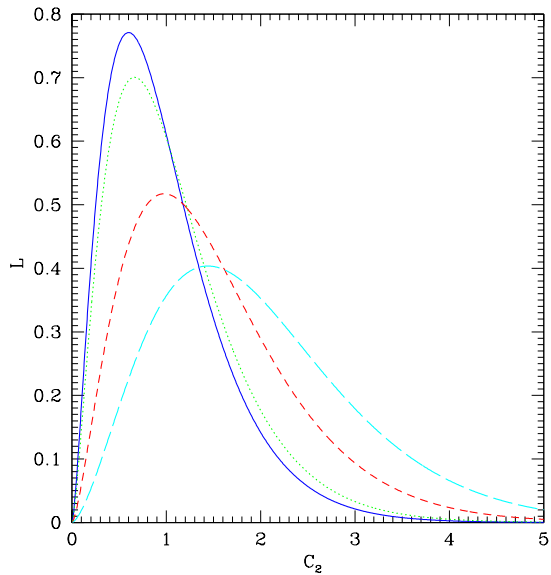}}}
\caption{\label{fig:like2} Quadrupole likelihood function for $\alpha=0$ (solid line), $0.1$ (dotted line), $0.5$ (short-dashed line) and $1.0$ (long-dashed line). The value of $C_2^{\rm I}$ has been arbitrarily set to unity.}
\end{figure}

It is clear that the effect of the anisotropic equation of state is to modify the temperature quadrupole so that the multipole coefficients are given by  
\begin{eqnarray}
a_{2,2}&=&a_{2,2}^{\rm I}-\sqrt{2\pi\over 15}J(\Omega_{\rm m},w)(\Delta w_1-\Delta w_2)\,,\nonumber\\
a_{2,1}&=&a_{2,1}^{\rm I}\,,\nonumber\\
a_{2,0}&=&a_{2,0}^{\rm I}+\sqrt{4\pi\over 5} J(\Omega_{\rm m},w)(\Delta w_1+\Delta w_2)\,,\nonumber\\
a_{2,-1}&=&a_{2,-1}^{\rm I}\,,\nonumber\\
a_{2,-2}&=&a_{2,-2}^{\rm I}-\sqrt{2\pi\over 15}J(\Omega_{\rm m},w)(\Delta w_1-\Delta w_2)\,,
\end{eqnarray}
where $a_{2m}^{I}$ is the stochastic quadrupole generated by initial metric perturbations, which need not have a diagonal covariance matrix (for example, see ref.~\cite{Battye:2009ze}). Using this, and assuming the two components are uncorrelated, we can compute $C_2$, the quadrupole coefficient of the angular power spectrum, whose mean value will be $C_2=C_2^{\rm I} + C_2^{\rm A}$. The {\em deterministic} power due to the anisotropic dark energy is
\begin{equation}
C_2^{\rm A}={8\pi\over 75}[J(\Omega_{\rm m},w)]^2(\Delta w)^2\,,
\end{equation}
with $(\Delta w)^2={\Delta w_{i}}^{j}{\Delta w_{j}}^{i}=2(\Delta w_1^2+\Delta w_2^2+\Delta w_1\Delta w_2)$. Note that the effect on the average power spectrum is additive. Rather than having the usual $\chi^2$ distribution with 5 degrees of freedom, the quadrupole likelihood now has a non-central $\chi^2$ distribution. Writing the ratio of power between the deterministic and stochastic components as $\alpha = C_2^{\rm A} /C_2^{\rm I}$, the variance of $C_2$ is given by
\begin{equation}
\left( \Delta C_{2} \right)^2 =  \frac{2 \left( 1 + 2 \alpha \right) }{5 \left(1+\alpha \right)^2} C_2^2\,.
\end{equation}
In Fig.~\ref{fig:like2} we illustrate the quadrupole likelihood function for various non-zero $\alpha$. In summary, the expectation value of $C_2$ will increase and the {\em fractional} cosmic variance will {\em decrease} for a non-zero deterministic  component.

\begin{figure}
\centering
\mbox{\resizebox{0.45\textwidth}{!}{\includegraphics[angle=0]{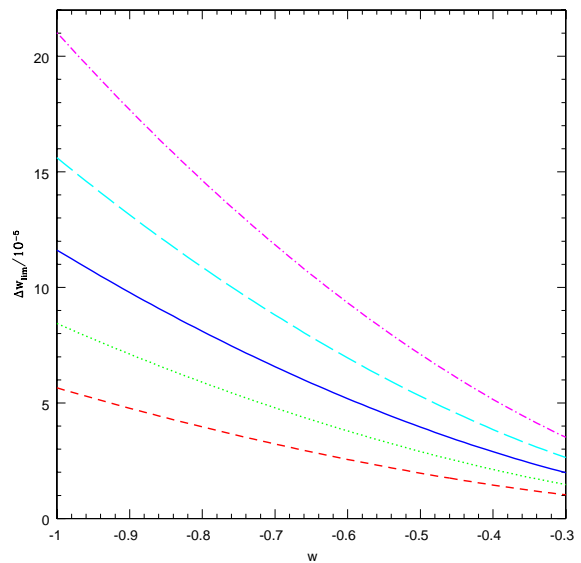}}}
\caption{\label{fig:constraint} The $2\sigma$ upper limit on the anisotropy, $\Delta w_{\rm lim}$, against $w$ for $\Omega_{\rm m}=0.1$ (dashed line), 0.2 (dotted line), 0.3 (solid line), 0.4 (short-dashed line) and 0.5 (dotted-short dashed line). }
\end{figure}

{\em Limits on the anisotropy of dark energy:} One can obtain a conservative constraint on $\Delta w$ by assuming that the entire quadrupole amplitude comes from the anisotropic effect. The result from WMAP5 is that $C_2<4.0\times 10^{-10}$ at 95\% confidence~\cite{Dunkley:2008ie,Nolta:2008ih}, which implies a constraint of $|\Delta w|<1.1\times 10^{-4}$ for  $w=-1$,  $6.0\times 10^{-5}$  for $w=-2/3$ and $2.2\times 10^{-5}$ for $w=-1/3$, all for $\Omega_{\rm m}=0.3$. The upper limit, $\Delta w_{\rm lim}=|\Delta w|$, for other values of $w$ and $\Omega_{\rm m}$ is presented in Fig.~\ref{fig:constraint}. Making the conservative assumption that $\Omega_{\rm m}<0.5$ we deduce that $|\Delta w|< 2.1\times 10^{-4}$. These limits are significantly stronger than those which come from type Ia SNe~\cite{Koivisto:2008ig,Cooke:2009ws} which are typically $|\Delta w|\sim 0.1$.

{\em Conclusions:} We have found that the consequences of  a time-dependent, but spatially homogeneous, anisotropic stress can be evolved in an anisotropic universe to first order. If we attribute this anisotropic stress to anisotropy in the dark energy which only comes to dominate a late times, then we can deduce strong limits on the anisotropy of the equation of state of the dark energy. Quantifying this in terms of $\Delta w$ we find that $|\Delta w|<2.1\times 10^{-4}$ for $\Omega_{\rm m}<0.5$

{\em Previous work on anisotropic universes:} We note that Campanelli {et al} have taken a similar approach in order to constrain the ellipticity of the surface of last scattering. In their calculation they introduce, {\it ad hoc},  the anisotropic stress associated with an elliptical surface of last scattering, $\sigma_{ij}=-e_{\rm rec}^2\delta_{i3}\delta_{j3}$ and constrain $e_{\rm rec}<10^{-2}$. Our calculation is similar in approach, but the anisotropic stress in our model is dynamical and calculated from the Einstein and conservation equations.

Our calculations of the sample variance are somewhat at odds with those presented in refs. ~\cite{Campanelli:2006vb,Campanelli:2007qn}. There, they found a non-zero anisotropic stress caused by an elliptical Universe could decrease the value of the observed quadrupole. Their approach was different, however, in that for each realization of the stochastic component they minimized the value of $C_2$ by changing the orientation of the ellipsoid. This is equivalent to assuming that the two components are correlated.  In our case, we assume the two components are uncorrelated, which leads to the increased expectation of $C_2$. It could be that there is correlation between the two components in our model generated at second order and this is presently being investigated. It is, however, likely to be a sub-dominant effect.

%----------------- ACKNOWLEDGMENTS -----------------------

%\section*{Acknowledgments} 
{\em Acknowledgments:} This research was supported by the Natural Sciences and Engineering Research Council of Canada. We thank Kris Sigurdson, Douglas Scott and Jim Zibin for useful discussions.

\end{document}